\documentclass[a4paper]{jpconf}
\usepackage{graphicx}
\usepackage{amssymb}

\newcommand {\snn}	{\sqrt{s_{_{\rm NN}}}}

\newcommand {\mpt}	{\langle p_{\rm T} \rangle}

\newcommand {\pT}       {p_{\rm T}}
\newcommand {\pttrig}	{p_{\rm T}^{\rm trig}}
\newcommand {\ptasso}	{p_{\rm T}^{\rm assoc}}

\newcommand {\Ettrig}	{E_{\rm T}^{\rm trig}}

\newcommand {\deta}	{\Delta\eta}
\newcommand {\dphi}	{\Delta\phi}

\begin{document}
\bibliographystyle{unsrt}

%
%
%
%
%
\title{Recent high $\pT$ measurements in STAR}

\author{Andr\'e Mischke \it{for the STAR Collaboration}}

\address{NIKHEF, Amsterdam, The Netherlands, \\
present address: Department of Subatomic Physics, Utrecht University, Princetonplein 5,
3584 CC Utrecht, The Netherlands.}

\ead{a.mischke@phys.uu.nl}

\begin{abstract}
%
After five years of data taking, the STAR experiment at the
Relativistic Heavy Ion Collider (RHIC) at Brookhaven National Laboratory
provides precise measurements of particle production at high transverse
momentum in p-p, d-Au, and Au-Au collisions at $\snn$~= 200~GeV. 
%
%
We review recent results on the flavor dependence of high $\pT$ particle
suppression and hadron particle spectra at $\snn$~= 62.4~GeV. 
New results on two-particle angular correlations for identified trigger
particles and for low momentum associated charged hadrons in p-p
and Au-Au as well as near-side $\Delta\eta$ correlations will be presented
and discussed.
\end{abstract}

\section{Introduction}

The RHIC facility at Brookhaven National Laboratory provides Au-Au
collisions at the highest energy presently available of $\snn$~= 200~GeV. 
The high
center-of-mass energy opens up the hard scattering regime which is 
accessed by the measurements of particle production at large transverse 
momentum and where pQCD is applicable. 
These high momentum particles originate from parton fragmentation 
in the early stage of the collisions. The scattered partons can be used 
to probe the produced medium of strongly interacting 
matter~\cite{JetQuen90,JetQuen92}. 

A significant suppression of high $\pT$ hadron production relative to a simple 
binary collision scaling from proton-proton collisions has been observed in 
central Au-Au collisions at RHIC~\cite{SuppPart03,SuppPart02}.
Additionally, it was found that jetlike correlations opposite to trigger jets 
disappear and that the elliptic anisotropy in hadron emission is large and
saturates at very high transverse momenta~\cite{BBCorr03,Flow03,AziAnCorr04}.
All these findings are consistent with the picture of parton energy loss (induced 
gluon radiation) of the scattered partons in the extremely dense medium
(jet quenching).
In contrast, no suppression effects were seen in d-Au collisions~\cite{dAuStar03}, 
which provide an important control measurement for the effects in cold nuclear 
matter. From the d-Au measurements it was concluded that the observations
made in Au-Au are due to final state interactions in the high density medium 
produced in such collisions. 

In this paper, recent high $\pT$ results from the STAR experiment~\cite{NimSTAR03}
are presented. From the observations so far, particle production can be described
dividing the transverse momentum axis into three ranges. The first range,
$0\lesssim\pT\lesssim2$~GeV/$c$, particles are produced via soft processes. 
These particles are usually called bulk matter. 
In the intermediate range, $2\lesssim\pT\lesssim6$~GeV/$c$, 
the interplay between the probe and the medium can be studied. Initial state
nuclear effects (e.g. Cronin) also contribute significantly in this range.
The $\pT$ range above $\sim6$~GeV/c seems to be the "clean" jet fragmentation 
regime where pQCD processes become dominant. Inclusive hadron spectra from p-p 
interactions are well described by pQCD.
In the next sections, recent measurements of the STAR collaboration concerning
inclusive particle spectra and particle correlations in Au-Au collisions at
200 and 62.4~GeV will be discussed using this classification.

\section{Low transverse momentum range}

\begin{figure}[t]
 \begin{minipage}{18pc}
  \hspace{-4mm}  
  \includegraphics[width=18pc,height=19pc]{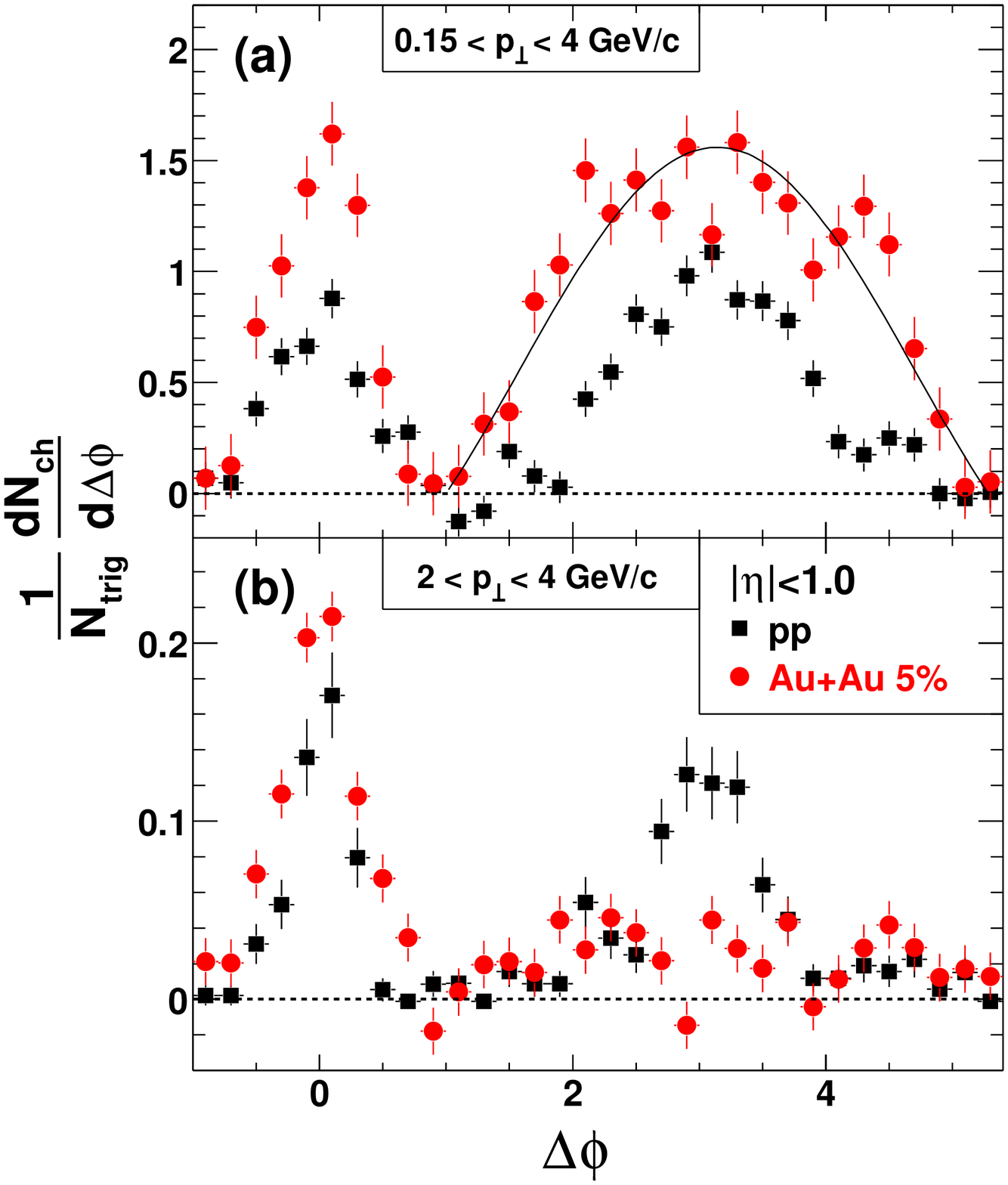}
  \caption{\label{fig1a} \footnotesize
Background subtracted dihadron azimuthal angular distribution
in p-p (solid circles) and 5\% most central Au-Au collisions (solid 
squares) 
for \mbox{$4<\pttrig<6$~GeV/$c$} and two associated $\pT$ ranges 
\mbox{0.15~$<\ptasso<4$~GeV/$c$} (\mbox{$2<\ptasso<4$~GeV/$c$}) 
upper (lower) plot ($|\eta|<1$). 
The curve in the upper plot shows the shape of an $[A-B\cos(\dphi)]$ function.}
 \end{minipage}
	\hspace{2pc}%
 \begin{minipage}{18pc}
 \vspace{-2mm}  
 \includegraphics[width=16pc]{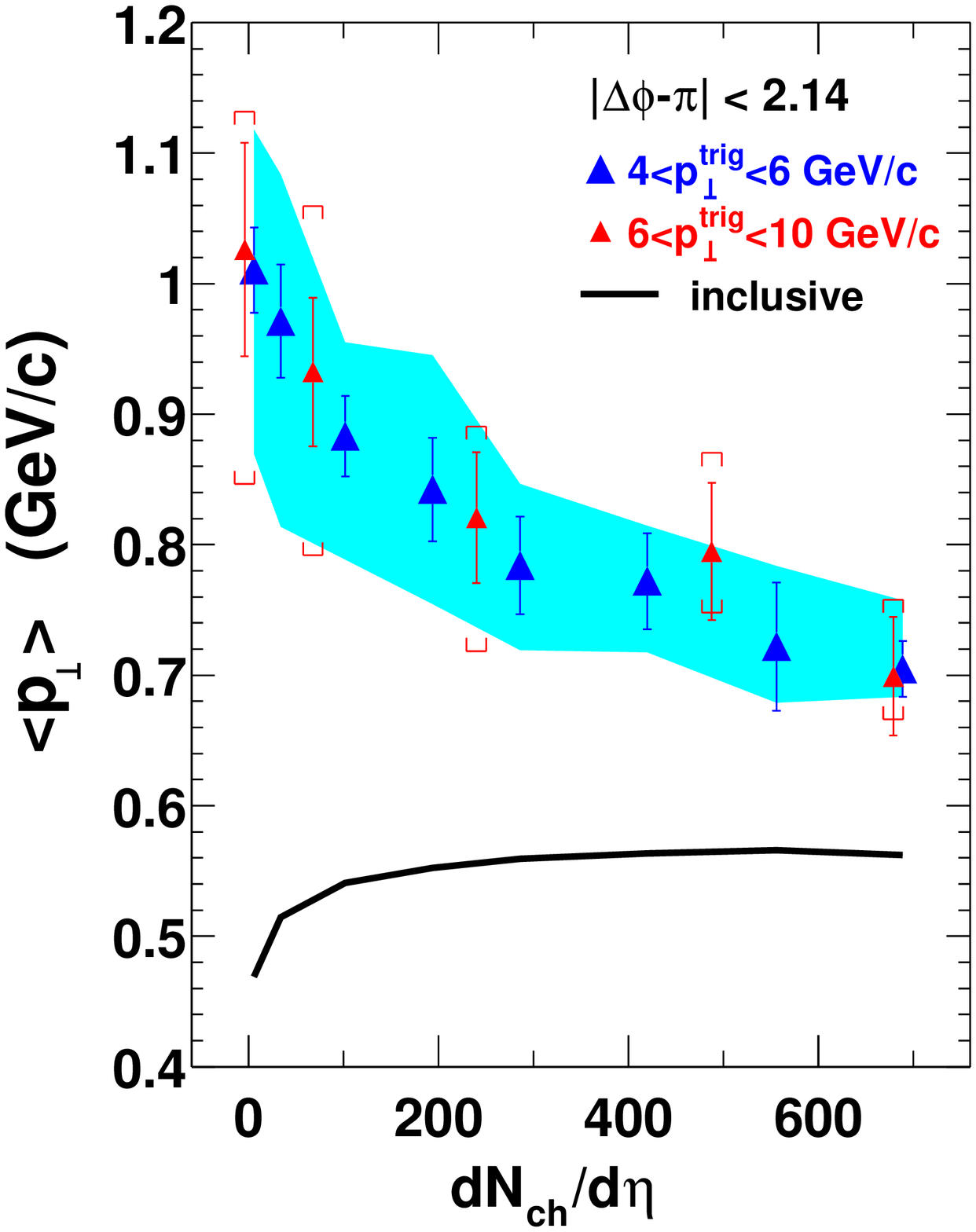}
 \vspace*{5mm}  
  \caption{\label{fig1b} \footnotesize
The $\mpt$ for the away-side associated charged hadrons selected in 
\mbox{4$<\pttrig<6$~GeV/$c$} (systematic errors are indicated by the 
shaded band) and \mbox{$6<\pttrig<10$~GeV/$c$} (systematic errors in 
caps) in Au-Au collisions. The leftmost data points are from p-p 
reaction. The solid line indicates the $\mpt$ for inclusive hadron 
production.}
 \end{minipage} 
\end{figure}

Although the subject of this paper is high $\pT$, low transverse 
momentum particle production is discussed for parton fragmentation 
products appearing in this kinematical range.
Due to full azimuthal coverage, the STAR detector has good 
capabilities to study leading particle correlations. The correlation
contribution from azimuthal anisotropy (elliptic flow) is measured
with high precision~\cite{Flow03,AziAnCorr04}.
It was observed that jetlike azimuthal angular ($\dphi$) correlations 
opposite to trigger jets in the
$\pT$ range $4<\pttrig<6$~GeV/$c$ are suppressed for associate particle
selected in $2<\ptasso<4$~GeV/$c$ in central Au-Au collisions,
whereas no suppression was measured in p-p, d-Au and peripheral 
collisions~\cite{dAuStar03}.
The method of leading particle correlation most likely favors the 
selection of high $\pT$ trigger particles near the medium surface. 
Therefore, the back-to-back parton has to go through the produced 
medium and can probe it. The suppression exhibits a continuous centrality 
dependence~\cite{BBCorr03}. Dihadron $\dphi$ correlation studies with 
respect to the reaction plane in the 20--60\% most central collisions 
have shown~\cite{AziAnCorr04} that the suppression depends on the path 
length traversed, which is in line with the jet-quenching scenario.

Because of momentum conservation the fragmentation products 
of the away-side jet must balance the momentum of the trigger jet. 
In Figure~\ref{fig1a}, upper panel,
the background (flow plus pedestal) subtracted $\dphi$ distribution 
for p-p and the 5\% most central Au-Au collisions are shown for the 
same trigger $\pT$, $4<\pttrig<6$~GeV/$c$, but for an associate $\pT$ 
range of $0.15<\ptasso<4$~GeV/$c$~\cite{LowPtRecoil05}. 
One observes a clear back-to-back correlation peak which is significantly 
broadened and shows a momentum balance shape. 
Similar results have been found by the NA45/CERES collaboration for 
pions with $\pT>1.2$~GeV/$c$ at CERN-SPS energies~\cite{NA45Corr04}.
Recently, it was suggested~\cite{Stoe05, MachCone04} 
that a double-peak structure can be expected from in-medium conical 
flow, caused by the energy deposition of partons traversing 
the produced matter faster than the speed of sound in the medium.
Similar theoretical calculations for Mach shock waves in nuclear reactions 
have been performed by~\cite{Hof76}.
The resulting shock wave may give rise to increased particle production 
at well-defined angles with respect to the parton direction 
($\dphi_{\rm flow\hspace{1mm} max.}\approx\pi\pm1.$).
Further analysis with higher statistics is needed to clarify the 
significance of such a away-side structure.

In addition, it was found that the mean $\pT$ of the back-to-back
associated charged hadrons in the trigger $\pT$ ranges
$4<\pttrig<6$~GeV/$c$ and $6<\pttrig<10$~GeV/$c$
decreases with centrality in Au-Au collisions as illustrated in
Figure~\ref{fig1b}. The solid line in the figure reflects the $\mpt$ for
inclusive hadron production. For the most central events, the associate 
particles approach this line indicating thermal equilibration of the 
away-side fragmentation products with the bulk medium. 

\section{Intermediate transverse momentum range}

\begin{figure}[t]
 \begin{minipage}{18pc}
  \hspace{-6mm}  
  \includegraphics[width=21pc,height=15pc]{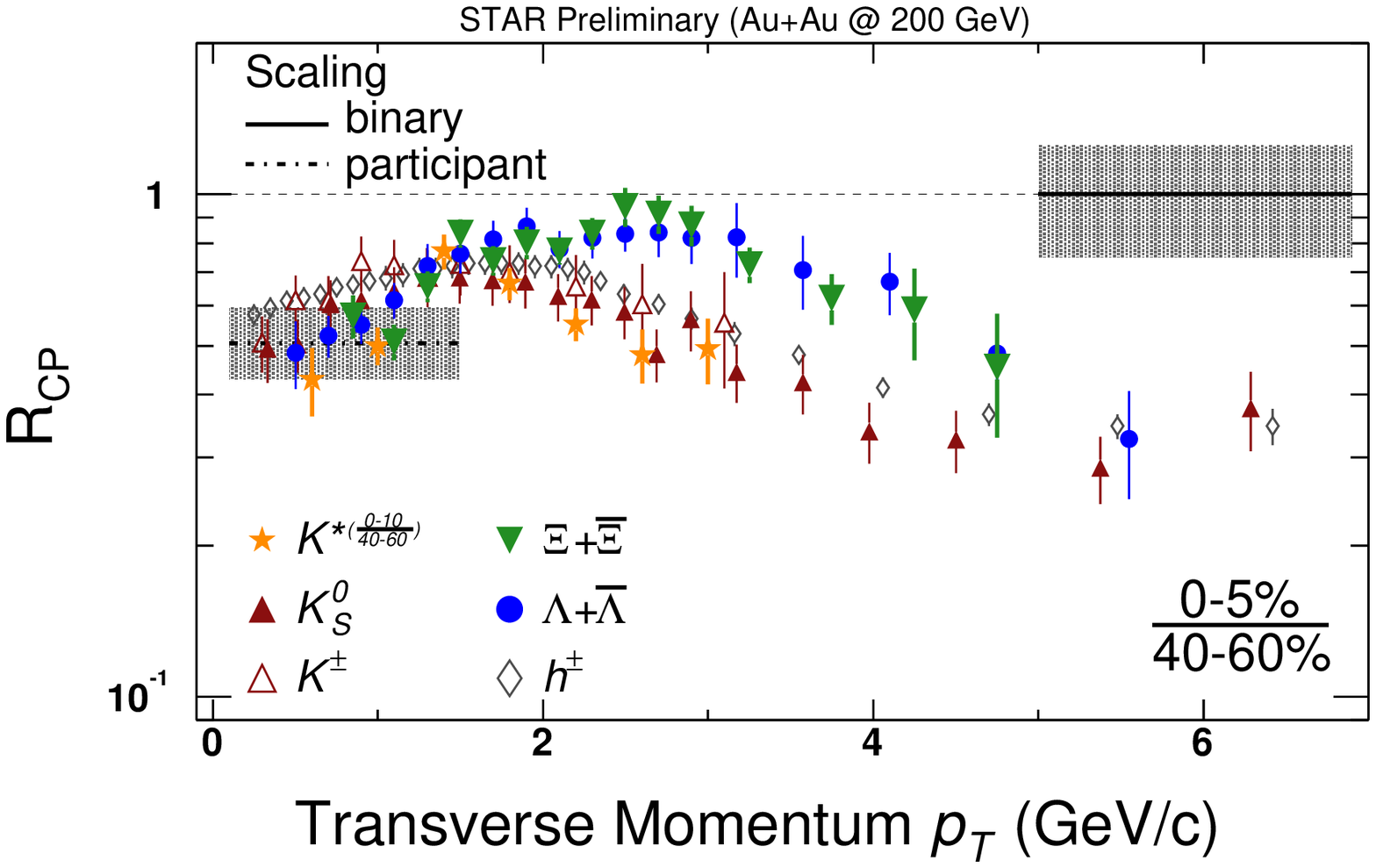}
  \caption{\label{fig2a} \footnotesize
Nuclear modification factor $R_{CP}(\pT)$ for
strange mesons and baryons using the centrality intervals 0--5\%
vs. 40--60\% of the Au-Au collision cross section at $\snn$~= 200~GeV.}
 \end{minipage}
  \hspace*{1pc}
 \begin{minipage}{18pc}
  \vspace*{-4mm}  
  \includegraphics[width=20pc]{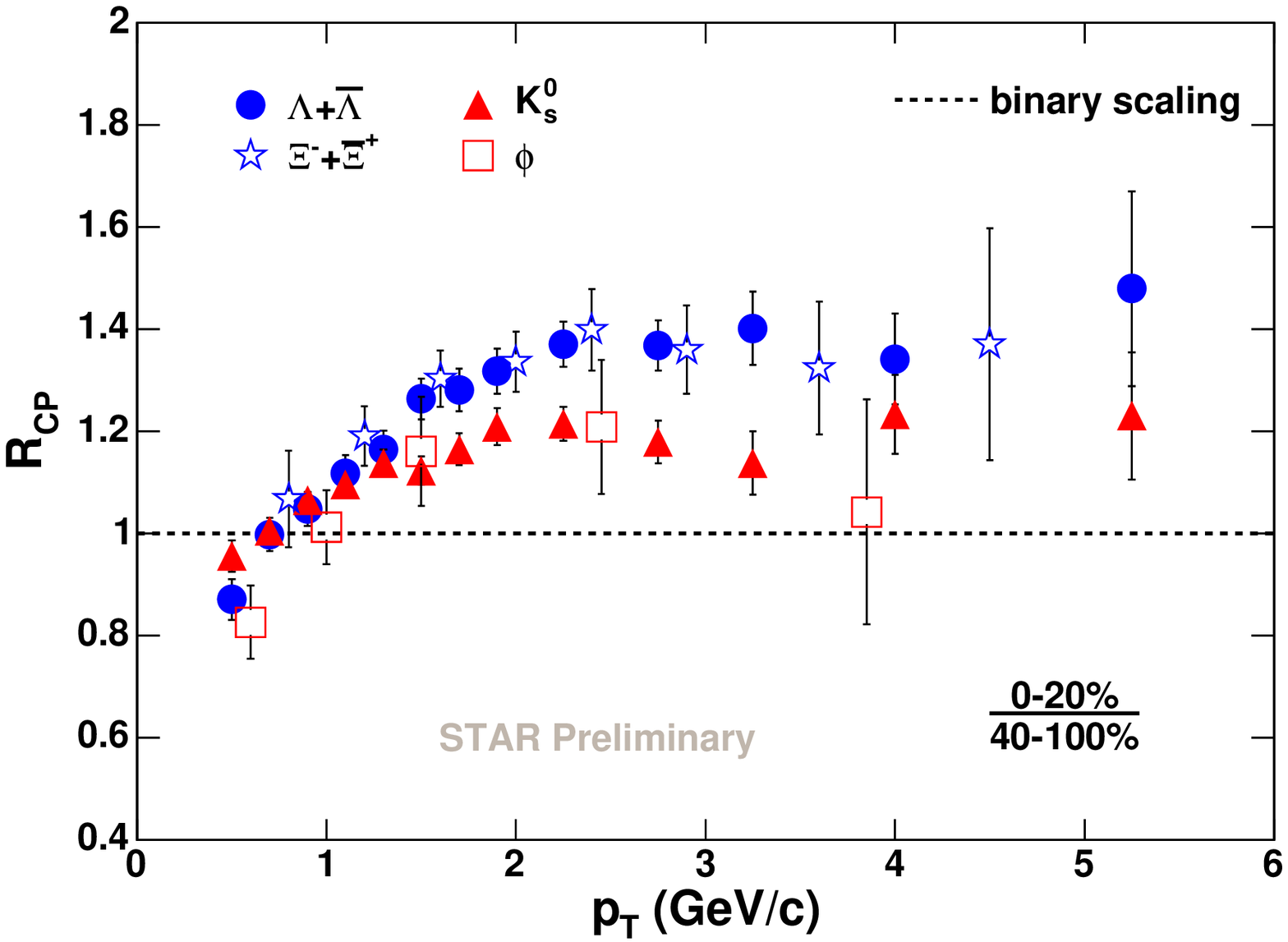}
  \caption{\label{fig2b} \footnotesize
$R_{CP}(\pT)$ for hyperons ($\Lambda$, $\Xi$), K$^{0}_{\rm s}$ 
and $\phi$ mesons in d-Au collisions. Here, 
the ratio is calculated using the centrality intervals 0--20\% vs. 
40--100\%.}
 \end{minipage} 
\end{figure}

Inclusive hadron suppression is quantified using the nuclear 
modification factor $R_{CP}(\pT)$, the ratio of the inclusive 
yields in central to peripheral collisions scaled by the number of 
binary collisions to take into account geometric effects. $R_{CP}$ 
is expected to be one for pointlike incoherent processes (absence of 
nuclear effects). 
Deviations from unity indicate nuclear modifications such as the 
Cronin effect, shadowing and modifications of the fragmentation 
function in the produced medium.

In central Au-Au collisions, a suppression of high $\pT$ particle
production is observed. Moreover, $R_{CP}$ exhibits a significant
meson-baryon pattern in the intermediate $\pT$ range 
($2\lesssim\pT\lesssim6$~GeV/$c$)~\cite{SuppPartSpez04,SuppPartSpezLam04}.
This pattern is plotted in Figure~\ref{fig2a} for charged hadrons,
kaons, hyperons ($\Lambda$, $\Xi$) and the K$^{\star}$ resonance. The 
$\phi$ follows the meson trend as well (not shown). 
This pattern suggests a collective hadron production mechanism such as
quark recombination or coalescence~\cite{RecCoalF04}.

$R_{CP}$ was also studied by the STAR collaboration for $\phi$ mesons
and hyperons in d-Au collisions~\cite{dAuHypPhi04} as illustrated in 
Figure~\ref{fig2b}.
Here, the yields are enhanced for $\pT>$~1~GeV/$c$ likely due to multiple
soft scatterings before the hard interaction (similar to the Cronin
effect). The meson-baryon pattern is also present with a much smaller
magnitude than in the Au-Au case ($\frac{\rm baryon}{\rm meson}\approx$~1.15). 
In d-Au collisions, 
the intermediate $\pT$ range should be dominated by jet fragmentation 
and one would expect one curve for all particles. Hence, the different 
behavior for mesons and baryons is a possible hint of the quark content
dependence of the Cronin effect (initial or final).
Future high statistic analysis have to show whether the curves for
mesons and baryons merge at even higher $\pT$ for d-Au and Au-Au.
For Au-Au collisions, one already notices a first hint around 6~GeV/$c$ 
(cf. Figure~\ref{fig2a}).

Remarkably, the integrated yields of the $\dphi$ correlation peaks for
identified K$^{0}_{\rm s}$ and $\Lambda$ as leading particle do not show
a significant flavor dependence at intermediate $\pT$~\cite{YGuo04}.
This result does not favor the thermal quark coalescence as the dominant
production source. Instead, it is more likely that one sees here the
interplay between parton fragmentation and the produced bulk matter.

\subsection{Pseudo-rapidity asymmetry in d-Au collisions}

STAR has performed systematic studies of the pseudo-rapidity asymmetry 
and the centrality dependence of charged hadron production in d-Au
collisions~\cite{EtaAsym04}.
One observes a significant $\eta$ dependence at the highest centrality
bin (0--20\% most central events) where the yields in the gold direction
are higher than in the deuteron direction.
The $\pT$ dependence of the back(gold)-to-forward(deuteron) asymmetry 
ratio is in qualitative agreement with the saturation 
model~\cite{Sat103,Sat203} but cannot be explained by initial state 
incoherent multiple scattering of the partons~\cite{XNWang03}.

\subsection{$\deta$ near-side dihadron correlations}

\begin{figure}[t]
 \begin{minipage}{18pc}
  \hspace{-2mm}  
  \includegraphics[width=19pc]{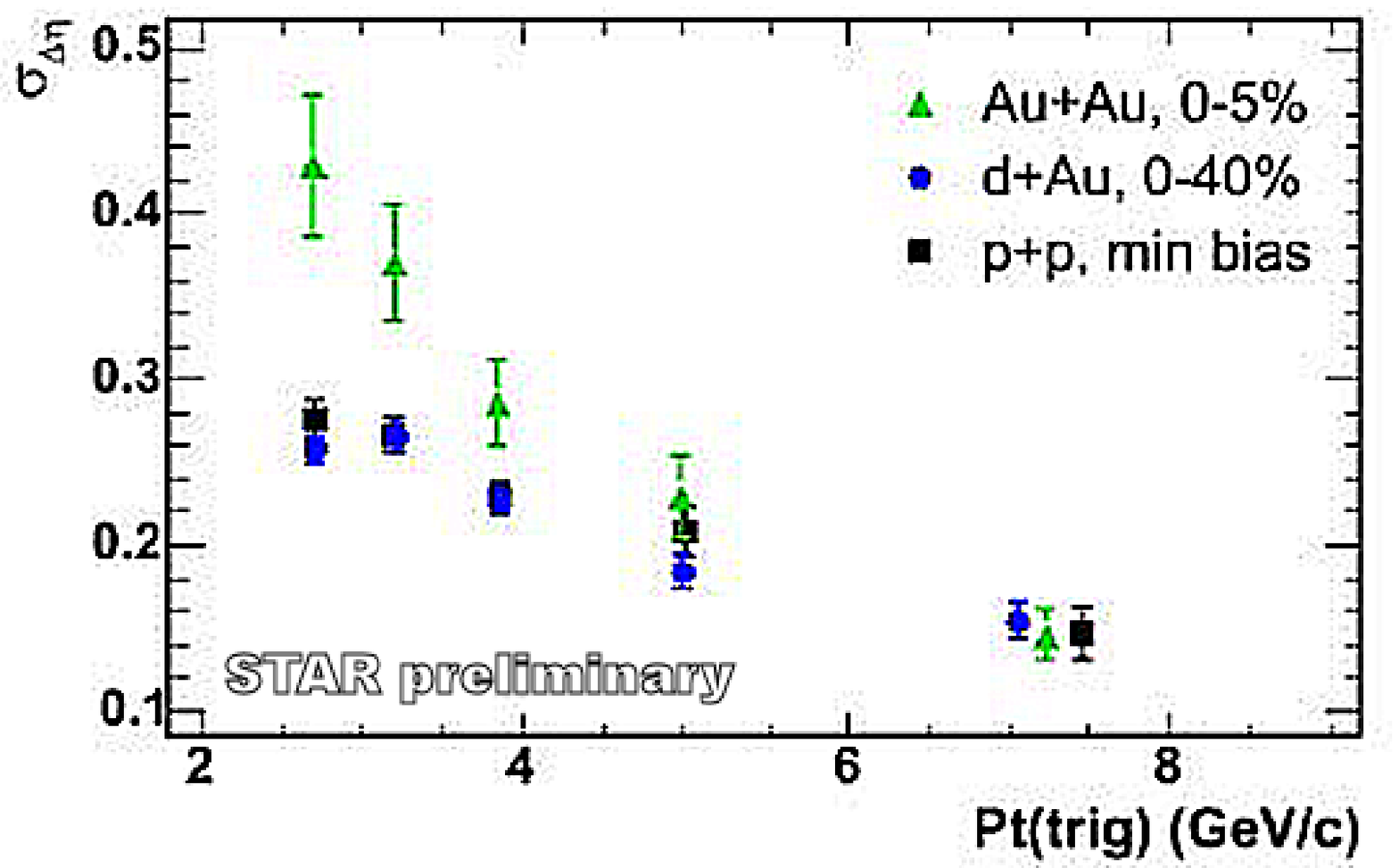}
  \caption{\label{fig3a} \footnotesize
Near-side $\deta$ correlation width of charged hadrons 
selected for 2~GeV/$c < \ptasso < \pttrig$ and $\dphi <$ 0.7
as a function of 
$\pttrig$ for p-p (square), d-Au (circle) and central Au-Au collisions 
(triangle).}
 \end{minipage}
	\hspace{0.5pc}
 \begin{minipage}{18pc}
  \includegraphics[width=19pc]{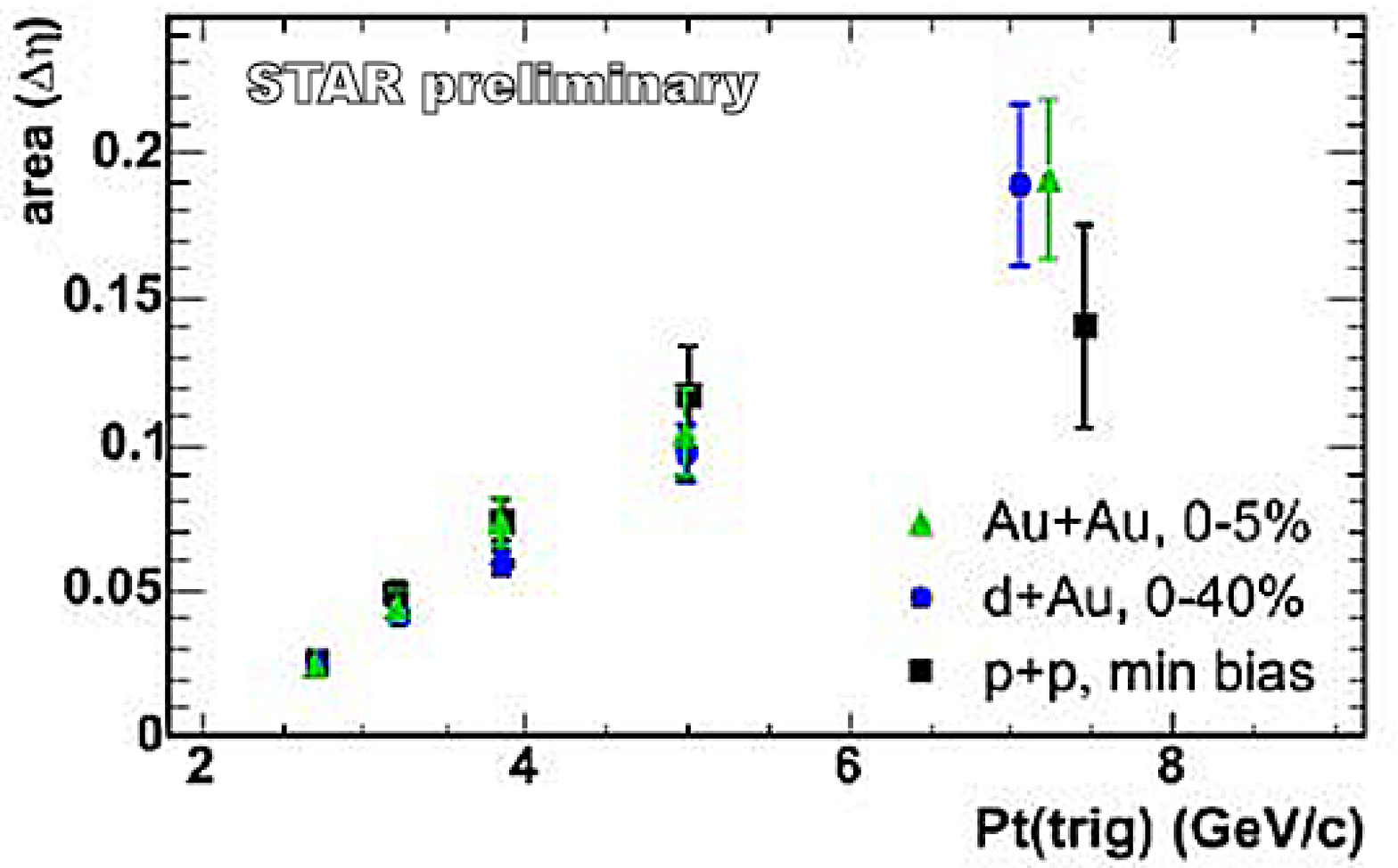}
  \caption{\label{fig3b} \footnotesize
Trigger $\pT$ dependence of the dihadron near-side $\deta$ correlation 
strength for p-p (square), \mbox{d-Au} (circle) and central Au-Au collisions.}
 \end{minipage} 
\end{figure}

Earlier STAR correlation analysis of soft hadrons ($\pT<$~2~GeV/$c$) in central
Au-Au collisions have shown~\cite{MiniJet04} that near-side correlations 
have a jetlike peak in $\dphi$ ($x-y$ plane) and $\deta$ ($y-z$ plane) with a 
similar width as measured in p-p reactions. This peak sits on top 
of a very broad, nearly flat distribution which can be attributed to long-range 
correlations~\cite{Fuq04,LowPtRecoil05}.
A new analysis~\cite{DanHP04} has been performed using high $\pT$ particles
which are originating from jet fragmentation. Figure~\ref{fig3a} shows the peak
width of the jetlike $\deta$ near-side correlation as a function of the trigger
$\pT$. This width is similar for p-p, d-Au and central Au-Au collisions at
$\pttrig>6$~GeV/$c$ but becomes broader for Au-Au when going to lower 
$\pttrig$. However, the correlation strength is similar for all three collisions 
systems as shown in Figure~\ref{fig3b}. These findings might likely be interpreted 
by the coupling of the medium-induced gluon radiation to the collective 
longitudinal flow~\cite{Armesto04,Volosh03}.

\subsection{High $\pT$ measurements at $\snn$~= 62.4~GeV}

\begin{figure}[t]
 \begin{minipage}{18pc}
  \hspace{-6mm}  
  \includegraphics[width=20pc]{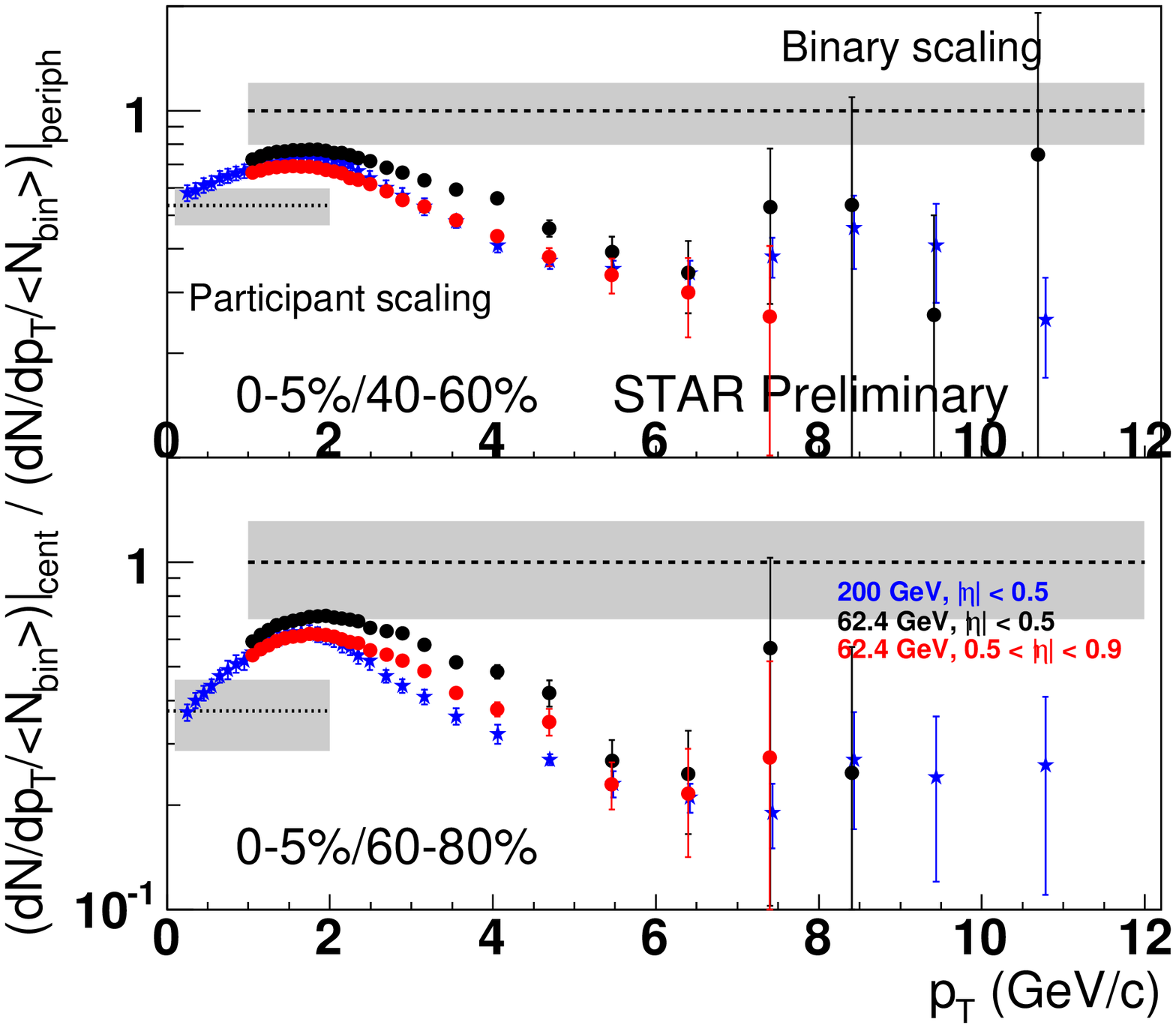}
  \caption{\label{fig4a} \footnotesize
Charged hadron nuclear modification factor $R_{CP}(\pT)$ at cms energy of 62.4 
(black and red points) and 200~GeV (blue points) for two different centrality 
interval ratios.}
 \end{minipage}
	\hspace{0.8pc}
 \begin{minipage}{19pc}
  \includegraphics[width=18pc]{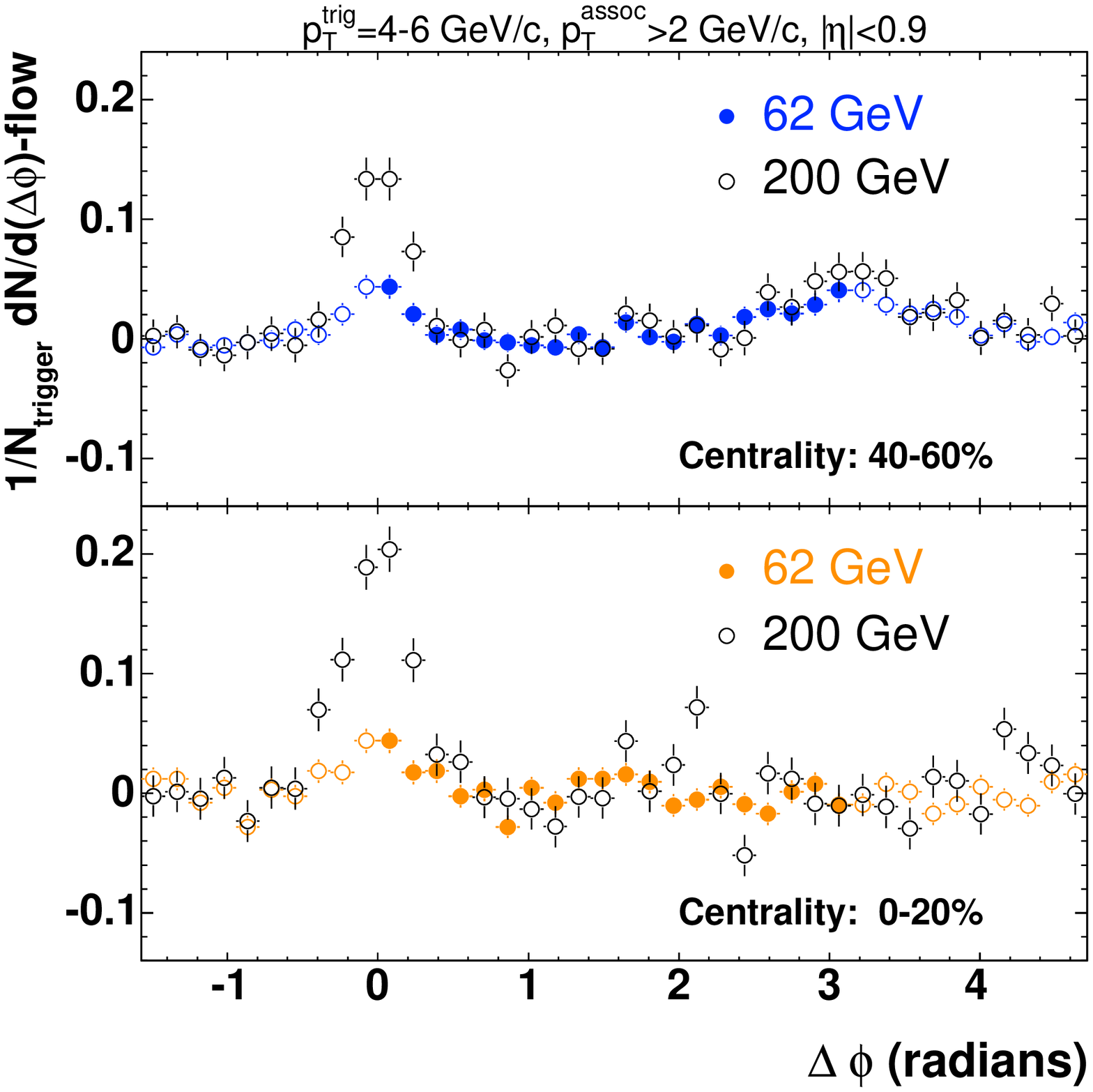}
  \caption{\label{fig4b} \footnotesize
Dihadron azimuthal angular correlation distribution at $\snn$~= 62.4 (full 
blue and orange points) and 200~GeV (open black points) for the kinematic 
cuts: \mbox{$4<\pttrig<6$~GeV,} $\ptasso>2$~GeV/$c$ and $|\eta|<0.9$.}
 \end{minipage} 
\end{figure}

High $\pT$ particle production is suppressed by a factor 
of about 5 in central Au-Au collisions. It has also been shown~\cite{DdEnt04}
that the nuclear modification factor is $R_{AA}\sim1$ at top CERN-SPS energy.
In 2004, the experiments at RHIC took Au-Au data at $\snn$~= 62.4~GeV to 
probe the onset of suppression. In the jet-quenching picture, one would
naively expect that the suppression is smaller at this energy because a
higher fraction of
quark jets are produced and the energy loss in the medium dependence on the 
color charged of the partons. At top RHIC energies, the number of quark and 
gluon jets should be about the same~\cite{JetQuenWang04}.
On the other hand, the smaller initial energy density leads to less parton 
energy loss whereas the steeper partonic $\pT$ spectrum generates larger 
leading hadron suppression for the same magnitude energy loss.

In Figure~\ref{fig4a}, the $R_{CP}(\pT)$ is shown for two different centrality
interval ratios for $\snn$~= 62.4 and 200~GeV. Within the errors, the
high $\pT$ suppression is similar above $\pT>$~6~GeV/$c$ for the two collision 
energies. Consequently, the suppression seems to be driven by the nuclear
geometry. The difference of the curves at the intermediate $\pT$ can be 
attributed to the Cronin effect.

Two-particle azimuthal angular correlation for two different centrality bins 
at 62.4 and 200~GeV is shown in Figure~\ref{fig4b}. In comparison to 200~GeV,
the near-side correlation strength is a factor $\sim3$ smaller at 62.4~GeV due
to the steeper underlying partonic spectrum~\cite{Fili05}. 
The away-side suppression is similar for both energies.

\section{High transverse momentum range}

\begin{figure}[t]
 \begin{minipage}{18pc}
  \hspace{-5mm}  
  \includegraphics[width=21pc,height=17pc]{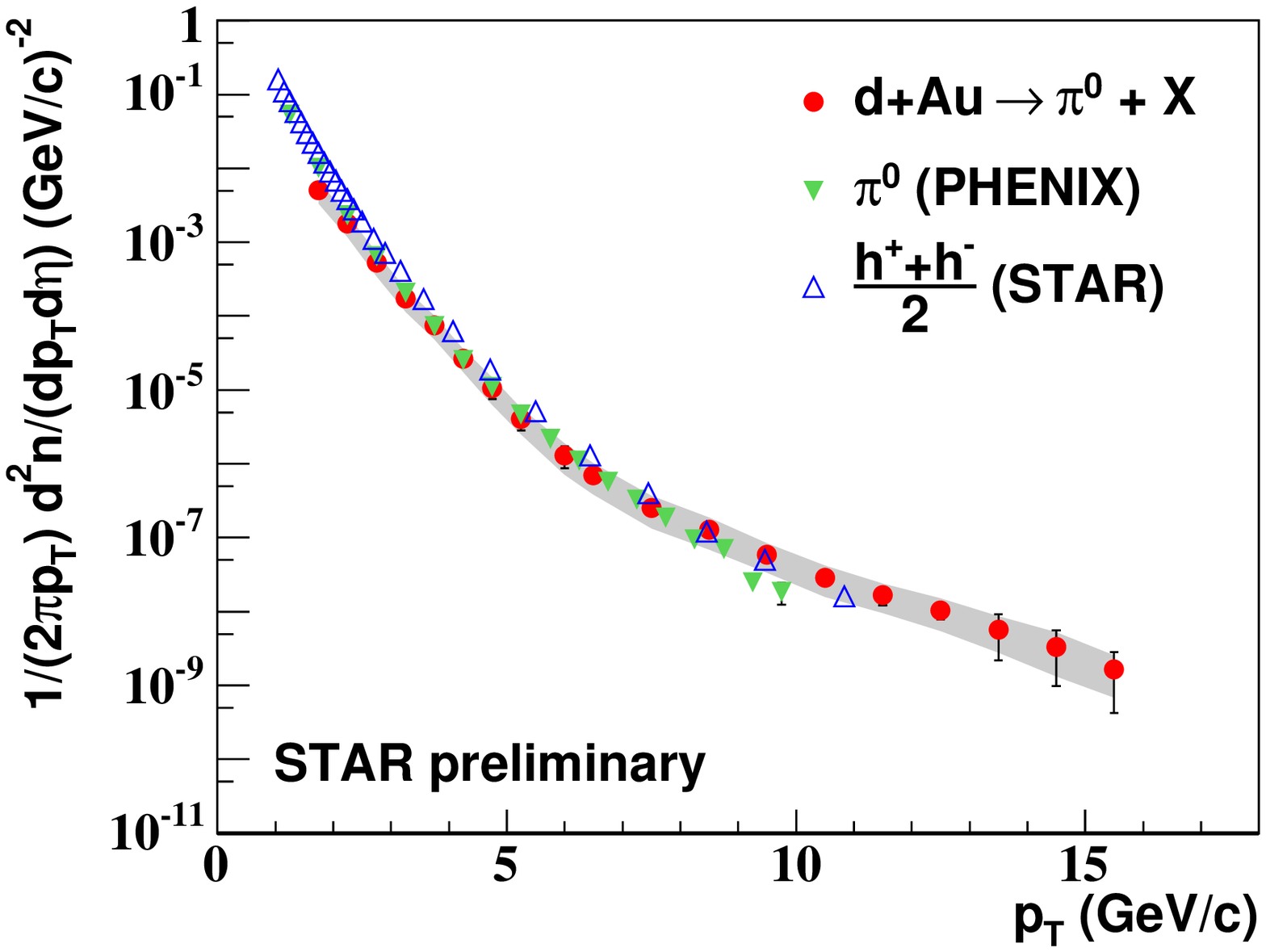}
  \caption{\label{fig5a} \footnotesize
Inclusive neutral pion spectrum in d-Au collisions at $0<\eta<1$ 
(full circles). The error bars (shaded band) represent the statistical (systematic) 
uncertainties. Previous STAR measurements of the charged hadron cross section
are shown by the open triangles, and the full triangles show $\pi^0$ results
from PHENIX.}
 \end{minipage}
	\hspace{0.8pc}
 \begin{minipage}{18pc}
  \includegraphics[width=18pc]{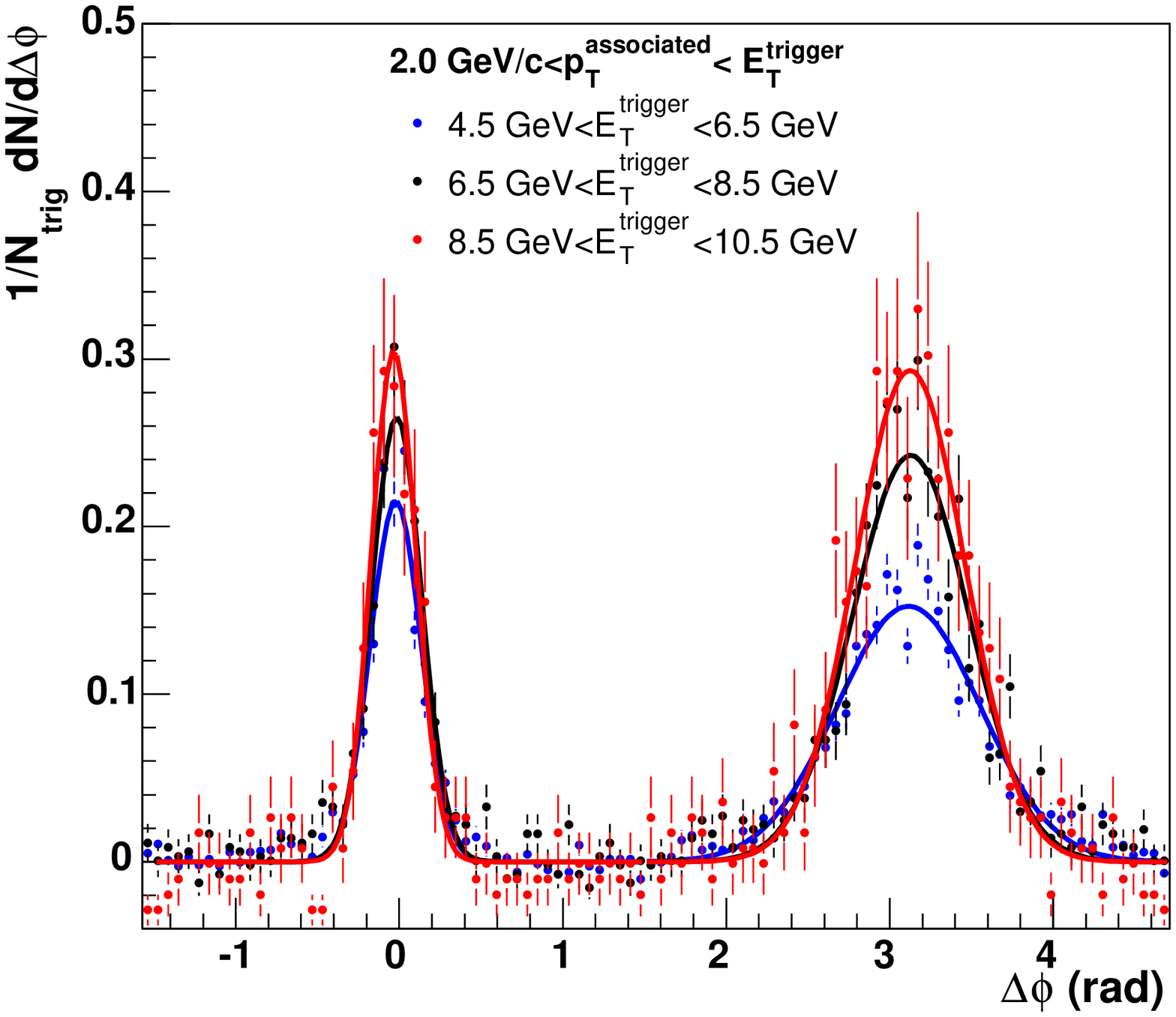}
  \caption{\label{fig5b} \footnotesize
Pedestal background corrected photon-charged hadron $\dphi$ correlation 
distribution for minbias d-Au collisions selected in the range 
\mbox{2.0~GeV/$c<\ptasso<\Ettrig$} and three different $\Ettrig$ ranges 
as indicated in the figure.}
 \end{minipage} 
\end{figure}

The large acceptance ($0\leq\phi\leq2\pi$ and $-1\leq\eta\leq1$) of the 
STAR Barrel Electromagnetic Calorimeter (BEMC)~\cite{NimEmc03,EtSTAR04} 
enables the study of identified particle spectra at even higher $\pT$ 
to probe where pQCD becomes dominant above nuclear and coalescence effects.
In addition, the high statistic 200~GeV Au-Au data set of Run IV will
certainly extend the inclusive hadron spectra to higher $\pT$.

Preliminary analysis of neutral pion production in d-Au collisions has 
shown the
capability of the BEMC~\cite{MisNPDC04}, with the $\pT$ spectrum 
measured up to $\sim15$~GeV/$c$ as plotted in Figure~\ref{fig5a}. 
The obtained spectrum shows reasonable agreement with previous STAR 
measurements on charged hadron cross sections~\cite{dAuStar03} and 
the PHENIX results on $\pi^{0}$ production~\cite{dAuPhenix03} as 
well as with NLO pQCD calculations~\cite{MisNPDC04}.
Figure~\ref{fig5b} shows the azimuthal angular photon - charged hadron 
correlation distribution in d-Au collisions at 200~GeV for different
$\Ettrig$ ranges~\cite{SubhQM04}. The leading photon can be selected up 
to $\sim11$~GeV/$c$. The obtained width and strength of the near- and 
away-side correlation peak are similar the the ones obtained from 
charged dihadron correlations~\cite{JanaCorr05}.

\section{Summary and future challenges}

In this paper, recent high $\pT$ results on inclusive particle spectra 
and two-particle $\dphi$ correlation at $\snn$~= 62.4 and 200~GeV
are presented. Similar behavior of the $R_{CP}$ and the dihadron 
correlation are observed which can be interpreted as the nuclear geometry
as the driving force of the suppression.
Correlation analysis with lower $\ptasso$ indicates thermal equilibration 
of the away-side fragmentation products with the bulk matter. 
The near-side $\deta$ correlation distribution can be interpreted as a 
coupling of the the parton energy loss to the collective flow.
d-Au collisions provide essential information for the deeper understanding
of the underlying mechanisms in Au-Au. The observed meson-baryon pattern
can not be exclusively attributed to quark coalescence.

Further studies and measurements are needed to understand the evolution of 
the jet-quenching effect~\cite{StarWhiteP05}. Minbias Cu-Cu collisions at 
$\snn$~= 62.4 and 200~GeV have been recorded this year. An additional low
energy 
Au-Au run will be necessary to map the energy dependence of the suppression 
effect between top SPS ($\snn$~= 17.4~GeV) and 62.4~GeV. More details on 
the modification of the fragmentation
function in the medium are needed which can be probed by direct photon
tagged correlation studies because the hard scattered photons are not
suffering from the surrounded medium. 
Heavy quark tagged correlations will offer the possibility to discriminate
between quark and gluon jets since heavier quark are dominantly produced
by gluon fragmentation.

\medskip
\section*{Acknowledgments}
\medskip
We thank the RHIC Operations Group and RCF at BNL, and the
NERSC Center at LBNL for their support. This work was supported
in part by the HENP Divisions of the Office of Science of the U.S.
DOE; the U.S. NSF; the BMBF of Germany; IN2P3, RA, RPL, and
EMN of France; EPSRC of the United Kingdom; FAPESP of Brazil;
the Russian Ministry of Science and Technology; the Ministry of
Education and the NNSFC of China; Grant Agency of the Czech Republic,
FOM and UU of the Netherlands, DAE, DST, and CSIR of the Government
of India; Swiss NSF; and the Polish State Committee for Scientific
Research.

\medskip
\section*{References}
\medskip
\bibliography{../../literatur_highpt.bib}

\end{document}